# Title: Utility of Multimodal Large Language Models in Analyzing Chest X-ray with Incomplete Contextual Information

*Authors*: Choonghan Kim[1], Seonhee Cho[1], Joo Heung Yoon[2]

*Author Affiliations*:

[1]Kim Jaechul Graduate School of AI, Korea Advanced Institute of Science & Technology (KAIST), Daejeon, Korea, Republic of

[2]Division of Pulmonary, Allergy, Critical Care, and Sleep Medicine, Department of Medicine, University of Pittsburgh, Pittsburgh, PA, United States

*Corresponding Author*: Joo Heung Yoon

*Address*: UPMC Montefiore, NW628 3459 Fifth Avenue Pittsburgh, PA 15213

*Email*: yoonjh@upmc.edu

*Phone*: +1-412-864-3199

*Fax*: +1-412-692-2260

*Funding*: Yoon JH has received grant support from NIH (National Institutes of Health) K23GM138984. This work was supported by Institute of Information & communications Technology Planning & Evaluation (IITP) grant funded by the Korea government (MSIT) (RS-2022-00143911, AI Excellence Global Innovative Leader Education Program)

**Abstract**

**Background**: Large language models (LLMs) are gaining use in clinical settings, but their performance can suffer with incomplete radiology reports. We tested whether multimodal LLMs (using text and images) could improve accuracy and understanding in chest radiography reports, making them more effective for clinical decision support.

**Purpose**: To assess the robustness of LLMs in generating accurate impressions from chest radiography reports using both incomplete data and multimodal data.

**Material and Methods**: We used 300 radiology image-report pairs from the MIMIC-CXR database. Three LLMs (OpenFlamingo, MedFlamingo, IDEFICS) were tested in both text-only and multimodal formats. Impressions were first generated from the full text, then tested by removing 20%, 50%, and 80% of the text. The impact of adding images was evaluated using chest x-rays, and model performance was compared using three metrics with statistical analysis.

**Results**: The text-only models (OpenFlamingo, MedFlamingo, IDEFICS) had similar performance (ROUGE-L: 0.39 vs. 0.21 vs. 0.21; F1RadGraph: 0.34 vs. 0.17 vs. 0.17; F1CheXbert: 0.53 vs. 0.40 vs. 0.40), with OpenFlamingo performing best on complete text (p<0.001). Performance declined with incomplete data across all models. However, adding images significantly boosted the performance of MedFlamingo and IDEFICS (p<0.001), equaling or surpassing OpenFlamingo, even with incomplete text.

**Conclusion**: LLMs may produce low-quality outputs with incomplete radiology data, but multimodal LLMs can improve reliability and support clinical decision-making.

## Introduction

Collecting and storing adequate clinical data is essential for any research, but errors can occur during this process. Often, the data available to researchers is incomplete or even misleading. Radiographic image reading error rates are around 3-5% (1), and the number had not changed much changed despite continuous efforts to systematically improve it (2). Even simple chart review showed the error rate in the electronic health record (EHR) could have been 9-10% (3).

The introduction of the artificial intelligence (AI) and more recently large language model (LLM) into medicine brought a wide array of fancy applications to numerous outstanding problems in current patient care and was deemed revolutionary to many sectors (4-7). LLMs' ability to handle complex unstructured data has led to applications that enhance patient experience, such as providing consultations (8), easing physician workloads through voice-recognized charting (9), and generating radiology reports (10).

However, some researchers argue that the rapid adoption of LLMs in medical research may be premature due to concerns about their performance in real-world clinical cases. Problems such as hallucinations (11), falsification (12), bias (13), and plagiarism (14) have been noted. These issues become more significant when LLMs are trained on incomplete or poor-quality data, resulting in biased and inaccurate outputs. This has raised concerns about the evaluation of LLMs trained on clinical data, including radiology reports (15), where objective evaluation methods remain insufficient.

We hypothesized that more corrupted text data would lead to worse LLM performance, highlighting the risks of using LLMs without knowing data integrity. Additionally, we expected that a multimodal LLM approach would improve text output quality compared to a unimodal (text-only) approach, emphasizing the value of a multifaceted strategy against incomplete data. We also anticipated that LLM-generated radiology reports could be objectively evaluated using statistical tools.

## Material and Methods

### Database

MIMIC-CXR, a deidentified, publicly available chest radiograph database collected from 2011 to 2016 was used for this analysis (16). MIMIC-CXR consists of 227,835 imaging studies from 65,379 patients, with each paired with radiologic reports (one pair of findings and impressions per image).

Three constraints were applied to exclude inappropriate data: 1) findings with fewer than three words, 2) findings shorter than impressions, and 3) findings with more than three masked words. We merged and filtered the dataset to remove the shortest and longest 25% of entries, resulting in training, validation, and test sets with 62,613, 1,000, and 1,000 entries, respectively. For time efficiency, 300 samples were randomly selected from the test set.

To test LLM robustness, random words or phrases were deleted from the findings at rates of 20%, 50%, and 80%, using a sub-word masking approach similar to BERT (17). (Fig 1) shows examples of these random text corruptions.

*Prompt Engineering*

We used few-shot learning (18) techniques to improve the performance of LLMs, since the performance of LLMs improves when given a few relevant examples even they were pre-trained. To select similar examples, we used BM25 (19) for text retrieval, which calculates relevance between a query and documents. We selected the top 2 examples based on BM25 scores, balancing memory constraints and optimizing computational costs, while adhering to the LLM's input limit of 2048 tokens.

*Large Language Models*

We chose multimodal models based on LLMs to test if multimodal data could improve performance. We identified models suitable for few-shot learning: OpenFlamingo (20), Med-Flamingo (21), and IDEFICS (22). These models were all based on the Flamingo (23) model, which handled sequences of image-text pairs using cross-attention layers called 'GATED XATTN-DENSE layers'.

OpenFlamingo is the publicly available version of Flamingo performing similarly on comparable tasks. Med-Flamingo is a vision-language model based on OpenFlamingo, pretrained with a medical dataset from textbooks. IDEFICS is also a vision-language model based on Flamingo, trained on the OBELICS (22) dataset with 353 million images and 115 billion text tokens. Summaries of these models are in (Supplementary Table 1).

*Evaluation of the LLM*

To evaluate the model, we used three common metrics for radiology reports, including ROUGE-L (24). ROUGE-L, which stands for Recall-Oriented Understudy for Gisting Evaluation, utilizes Longest Common Subsequence (LCS) statistics to measure text summary quality, providing a balanced F-score based on recall and precision.

Recall(R):

$$R = \frac{Length\ of\ LCS\ between\ candiate\ summary\ and\ reference\ summary}{Length\ of\ reference\ summary}$$

Precision ($P$):

$$P = \frac{Length\ of\ LCS\ between\ candiate\ summary\ and\ reference\ summary}{Length\ of\ candidate\ summary}$$

$F - \text{measure}(F)$:

$$F = \frac{2 \times R \times P}{R + P}$$

F1-RadGraph (25) is an F-score-based metric that measures the factual accuracy, coherence, and thoroughness of radiology reports compared to a reference. It uses PubMedBERT (26), fine-tuned on the RadGraph (27) dataset, to assess entities and their relationships. The metric assigns a score based on how well the entities and relationships in the generated output match those in the reference, representing these as a graph $G(V, E)$, where $V$ is the set of nodes (entities) and $E$ is the set of edges (relationships).

For the generated report $y$, this metric creates a new set of triplets $\overline{V_y}$ as follows:

$$\overline{V_y} = \{(v_i, v_{iL}, \epsilon_i)\}_{i=1}^{|V|}$$

Each triplet consists of an entity $v_i$, its corresponding label $v_{iL}$, and an indicator $\epsilon_i$ that shows whether the entity has a relation in $E$. $\epsilon_i$ is defined as:

$$\epsilon_i = \begin{cases} 1, & \text{if } v_i \text{ has a relation in } E \\ 0, & \text{otherwise} \end{cases}$$

It proceeds to construct the same set for the reference report $\hat{y}$ and denotes this set $\overline{V_{\hat{y}}}$. The harmonic mean of precision and recall between their respective sets $\overline{V_{\hat{y}}}$ and $\overline{V_y}$ is defined as:

$$F1 = \frac{2 \times \text{Precision} \times \text{Recall}}{\text{Precision} + \text{Recall}}$$

where:

$$\text{Precision} = \frac{\text{Number of correct elements in } \overline{V_y} \text{ that match } \overline{V_{\hat{y}}}}{\text{Total number of elements in } \overline{V_y}}$$

$$\text{Recall} = \frac{\text{Number of correct elements in } \overline{V_y} \text{ that match } \overline{V_{\hat{y}}}}{\text{Total number of elements in } \overline{V_{\hat{y}}}}$$

F1CheXbert (28) calculates the F1 score by comparing CheXbert (29)'s predictions with the reference report. Using a biomedically pretrained BERT, CheXbert predicts 14 observations, focusing on 5 key conditions: Cardiomegaly, Edema, Consolidation, Atelectasis, and Pleural Effusion. Overall performance was measured using the micro average for unbalanced datasets.

$$F1_{micro} = 2 \times \frac{P_{micro} \times R_{micro}}{P_{micro} + R_{micro}}$$

where:

$$P_{micro} = \frac{\sum_{i=1}^{n} \text{True Positives}_i}{\sum_{i=1}^{n} \text{Predicted Positives}_i}$$

$$R_{micro} = \frac{\sum_{i=1}^{n} \text{True Positives}_i}{\sum_{i=1}^{n} \text{Actual Positives}_i}$$

In this context, n represents the number of classes (Cardiomegaly, Edema, Consolidation, Atelectasis, and Pleural Effusion mentioned above), and True Positives, Predicted Positives, and Actual Positives are summed across all these classes. The overall workflows from data acquisition to performance evaluation is summarized in (Fig 2).

*Computational Cost of Large Language Models*

We calculated each model's parameters, computational budget (GPU hours), infrastructure used, LLM cost per token, and computing time. OpenFlamingo had 3 billion parameters, while MedFlamingo and IDEFICS had 9 billion each. These models used 30GB of GPU memory, increasing to 40GB during inference on a NVIDIA A6000 GPU. Budget estimates were compared to AWS (30).

*Statistical Analysis*

We used the scipy.stats library (31) in Python 3.11 for statistical analysis. P-values were calculated from ROUGE-L and F1RadGraph scores for 300 test samples. For F1CheXbert (micro average), we divided the samples into 10 groups and calculated micro averages. The Wilcoxon test was applied where normality wasn't met, with significance set at $p<0.05$. For the five diseases, we reported average performance without p-values due to limited samples. All analysis was performed in Python 3.11.

## Results

A total of 3,146 data labels (2.46%) were found to be incomplete and not usable for research. The remaining data included positive, negative, and uncertain labels for various conditions. In the training set, the most common conditions were 'No Finding' (47.6% positive), 'Lung Opacity' (15.5% positive), and 'Pleural Effusion' (12.4% positive). Other notable conditions included 'Atelectasis' (11.2% positive), 'Edema' (8.4% positive), and 'Cardiomegaly' (7.2% positive). The test set showed similar distributions, with 'No Finding' (47.5% positive), 'Lung Opacity' (15.1% positive), and 'Atelectasis' (11.4% positive). Other conditions included 'Pleural Effusion' (10.5% positive), 'Cardiomegaly' (7.6% positive), and 'Edema' (7.6% positive).

Using three metrics — ROUGE-L, F1RadGraph, and F1CheXbert — the text-only models of OpenFlamingo, MedFlamingo, and IDEFICS had average scores of 0.39, 0.21, and 0.21 using ROUGE-L, and 0.34, 0.17, and 0.17 with F1RadGraph, and 0.53, 0.40, and 0.40 for F1CheXbert with full text input, respectively (Table 1). Since MedFlamingo and IDEFICS used the same LLaMA 7b model, their results were identical.

When combining text data with images, the multimodal LLMs—OpenFlamingo, MedFlamingo, and IDEFICS— scored 0.40, 0.41, and 0.46 on ROUGE-L, 0.35, 0.40, and 0.44 on F1RadGraph, and 0.51, 0.61, and 0.62 on F1CheXbert, respectively (Table 1).

Med-Flamingo and IDEFICS outperformed their unimodal (text-only) versions in conditions with full findings, and with 0.2 and 0.5 levels of data corruption. This improvement was statistically significant across all three metrics (Fig. 3). OpenFlamingo performed better than its text-only version only at the 0.8 data corruption level, with significant differences in ROUGE-L and F1RadGraph metrics (Fig. 3).

(Table 2) showed the comparison between the fine-tuned and general models in a multimodal setting. Med-Flamingo, the clinically fine-tuned model, outperformed the generalist model, OpenFlamingo, in all but two cases: corrupted 0.8 on ROUGE-L and corrupted 0.8 on F1RadGraph.

For F1RadGraph, Med-Flamingo showed better performance than OpenFlamingo under full, 0.2, and 0.5 data corruption levels, with statistically significant differences (Supplementary Fig. 1). The fine-tuned model also performed better for the five representative disease categories compared to the general model (Supplementary Table 2). Correct or incorrect output samples of the fine-tuned multimodal LLM (MedFlamingo) were outlined for each of the five diseases (Fig. 4).

For computational cost, the processing times were as follows:

- OpenFlamingo: 27 seconds per sample; 2.25 hours.

- OpenFlamingo text-only: 38.44 seconds per sample; 3.20 hours.

- MedFlamingo: 13.80 seconds per sample; 1.15 hours.

- MedFlamingo text-only: 8.06 seconds per sample; 0.67 hours.

- IDEFICS: 1.81 seconds per sample; 0.15 hours; same performance for its text-only version as MedFlamingo LLM.

Using a similar GPU to ours (A100 with 80GB) on AWS, costing $40.96 per hour, the costs were:

- OpenFlamingo (multimodal) & OpenFlamingo text-only: $92.16 and $131.20

- MedFlamingo (multimodal) & MedFlamingo text-only: $47.01 and $27.50

- IDEFICS (multimodal) & IDEFICS text-only: $6.16 and $27.50, respectively.

In multiple comparisons, a fine-tuned multimodal LLM (MedFlamingo) or a highly functional LLM with an image decoder (IDEFICS) could achieve better efficiency compared to their unimodal counterparts.

## Discussion

In this study, we found the yield and accuracy of the 3 different LLMs to be decreased proportionately with the increasing corruption of input text data from a radiography report. In addition, we identified the superiority of the multimodal LLM approach including image processors combined with text data to unimodal LLM where only text data were utilized. In clinical literature, this is the first attempt to quantify the performance of various LLMs in setting of input data corruption, apply multimodal approach to improve the performance of those LLMs, and statistically compare the performance between those model pairs in radiologic data with showcasing cost-effectiveness of LLMs.

As much as the strong potential utility of LLM in healthcare research, there are growing concerns of the inadequate behaviors when the models are linked with important decision-making process or social determinants of health. When the LLM was used to create the resuscitation guideline with its semantic content generation capacity, the model was inaccurate and inconsistent with potential harmful effect in real-life use (32). A large difference in the estimation of race and gender distribution was found between the LLM-derived clinical conditions and their real prevalence, signifying that the general model training of the existing corpus of clinical information does not accurately reflect true condition and biased by the input data (33). Even though not in area of LLM, deep network could identify patients' race from chest X-ray images, which authors were not able to decipher underlying mechanisms (33). Across different topics and methodologies, these reports converge into the importance of the quality and adequacy of the training data and its influence on downstream models. Thus, the development of medical language models that rely solely on text data can carry potential risks. Developing medical multimodal foundation models that can simultaneously utilize both text and image data might reduce this chasm, as seen in current work.

The highly semantic representation of complex statistical relationship among different elements of healthcare data put LLM in a spotlight of current research methodology in clinical medicine. However, among a few other pitfalls, there are growing concerns in using LLM when input data are corrupted. The corrupted or damaged data would lead to suboptimal ML model performance (34) but very hard to avoid at the step of automated data de-identification (35). Our database also had such inherent corruption, as well as 'frameshift' issue in data format – which is also not uncommon in large-scale dataset. In our case, the frameshift issue was observed when impression column of the chest X-ray contained contents belonged to findings, or findings column of the chest X-ray report had impression of the X-ray and impression column was empty. Considering those incompleteness, incremental corruption on random data points was performed to address the relationship between the input data quality and the performance of the LLM. The outcome of our exercises highlighted the importance of quality control in LLM performance in setting of incomplete input data and suggested the option of multimodal LLM as a solution.

In the field of radiology, fine-tuning of LLM became a heated topic with its capacity to improve contextual understanding (36). The art of the prompt engineering in construction also could enhance the performance of LLM. However, we showed that blind fine-tuning would not improve the performance of the model if the input data were heavily corrupted. Fine-tuning is an effective way to infuse clinical knowledge into deep learning models (21, 37, 38). And high-quality data is crucial for fine-tuning training to improve model performance (39). However, we found that if a model was fine-tuned with corrupted data, it may not only reduce degradation but also show lower performance when exposed to uncorrupted data. As shown, our experiment can lead to developing a robust model that retains performance with the quality of input data.

Comparison between multimodal LLM and unimodal version indicates that multimodal LLMs significantly improve performance, especially with data corruption. Using multimodal LLM without additional clinical data training requires few-shot learning, which means showing similar examples to the model. Unlike text-only models, not many multimodal LLMs can handle few-shot learning. Many state-of-the-art multimodal LLMs, like BLIP-2 (40) and LLaVA (41), can only handle multiple text pairs with a single image. These inherent limitations are the reason that we used only Flamingo variants in our experiment. To develop a multimodal LLM that performs robustly under data corruption, it is necessary to consider multiple vision-text pairs.

The cost analysis highlighted LLM efficiency. Discussions on AI's cost-effectiveness in radiology (42, 43) led us to find that generating text with LLMs costs $40.96 per hour, lower than a radiologist's $115.00 per hour (44). This suggests potential cost savings, especially with a multimodal approach. Multimodal models like OpenFlamingo and IDEFICS were often cheaper and faster than unimodal versions, showing that well-optimized multimodal models can be more efficient. Further research could enhance their clinical use.

This study has several limitations. First, the data came from a single source of publicly available database so the overall trajectories or trends of performance of LLM in different environment might be hard to generalize. Second, the choice of LLM does not include some of the most efficient and well-known general models, such as GPT-4, LLaMA2, or MedPalm2. In our case, the working mechanism of main LLM (Flamingo) is analogous to the GPT-3 (45), so we assumed the overall sematic construction would not be vastly different. Also, the use of GPT-4 and MedPalm2 were limited due to their restricted API in an academic setting. Also, we used MedFlamingo, which was a state-of-the-art medical LLM for in-context learning that surpassed all other available LLMs as of September 2023 (21). With that, we think we used the most clinically relevant model for this study. Third, the multimodal approach that we selected was limited to images, which are raw radiographic data. Since there could be many different data types other than text data (structured data such as numeric, or unstructured data including physiologic waveforms), our results might not be fully generalized across different data types. Nevertheless, our approach could be still valid when confined to imaging interpretation tasks. Lastly, there could be other performance evaluation metrics for LLM to quantify the accuracy and semantic relationships.

In conclusion, to the best of our knowledge, this is the first comparison between multimodal LLM and text-only approaches over radiologic data especially when data integrity is suboptimal. To develop a truly practical LLM on highly noisy and healthcare data without guaranteed accuracy, multimodal approach should be considered rather than relying on text-only LLMs.

# References

1. Robinson PJ, Wilson D, Coral A, et al. Variation between experienced observers in the interpretation of accident and emergency radiographs. Br J Radiol 1999;72:323-30.

2. Brady AP. Error and discrepancy in radiology: inevitable or avoidable? Insights Imaging 2017;8:171-182.

3. Feng JE, Anoushiravani AA, Tesoriero PJ, et al. Transcription error rates in retrospective chart reviews. Orthopedics 2020;43:e404-e408.

4. Blinded for anonymity.

5. Janssen BV, Kazemier G, Besselink MG. The use of ChatGPT and other large language models in surgical science. BJS Open 2023;7:zrad032.

6. Betzler BK, Chen H, Cheng C-Y, et al. Large language models and their impact in ophthalmology. Lancet Digital Health 2023;5:e917-e924.

7. Romano MF, Shih LC, Paschalidis IC, et al. Large language models in neurology research and future practice. Neurology 2023;101:1058-1067.

8. Benary M, Wang XD, Schmidt M, et al. Leveraging large language models for decision support in personalized oncology. JAMA Network Open 2023;6:e2343689

9.	Krishnamoorthy R, Nagarajan V, Pour H, et al. Voice-Enabled Response Analysis Agent (VERAA): Leveraging Large Language Models to Map Voice Responses in SDoH Survey. AMIA Jt Summits Transl Sci Proc 2024 ;2024:258-265.

10.	Tie X, Shin M, Pirasteh A, et al. Personalized Impression Generation for PET Reports Using Large Language Models. J Imaging Inform Med 2024:1-18.

11.	Huang L, Yu W, Ma W, et al. A survey on hallucination in large language models: Principles, taxonomy, challenges, and open questions. arXiv preprint arXiv:2311.05232. 2023.

12.	McGowan A, Gui Y, Dobbs M, et al. ChatGPT and Bard exhibit spontaneous citation fabrication during psychiatry literature search. Psychiatry Res 2023;326:115334.

13.	Karabacak M, Margetis K. Embracing large language models for medical applications: opportunities and challenges. Cureus 2023;15(5).

14.	Pal S, Bhattacharya M, Islam MA, et al. AI-enabled ChatGPT or LLM: a new algorithm is required for plagiarism-free scientific writing. Int J Surg 2024;110:1329-1330.

15.	Doshi R, Amin KS, Khosla P, et al. Quantitative evaluation of large language models to streamline radiology report impressions: A multimodal retrospective analysis. Radiology 2024;310:e231593.

16. Johnson AE, Pollard TJ, Berkowitz SJ, et al. MIMIC-CXR, a de-identified publicly available database of chest radiographs with free-text reports. Sci Data 2019;6:317.

17. Devlin J, Chang M-W, Lee K, et al. Bert: Pre-training of deep bidirectional transformers for language understanding. In: Proceedings of the 2019 Conference of the North American Chapter of the Association for Computational Linguistics: Human Language Technologies; 2019 Jun 2-7; Minneapolis, MN. Stroudsburg, PA: Association for Computational Linguistics, 2019:4171-4186.

18. Brown T, Mann B, Ryder N, et al. Language models are few-shot learners. In: Advances in Neural Information Processing Systems 33 (NeurIPS 2020); 2020 Dec 6-12; Virtual. Red Hook, NY: Curran Associates, Inc., 2020:1877-1901..

19. Robertson S, Zaragoza H. The probabilistic relevance framework: BM25 and beyond. Found. Trends Inf. Retr 2009;3:333-89.

20. Awadalla A, Gao I, Gardner J, et al. Openflamingo: An open-source framework for training large autoregressive vision-language models. arXiv preprint arXiv:2308.01390. 2023.

21. Moor M, Huang Q, Wu S, et al. Med-flamingo: a multimodal medical few-shot learner. In: Proceedings of the 3rd Machine Learning for Health Symposium (ML4H); 2023 Dec 6-7; New Orleans, LA. PMLR, 2023:353-367

22. Laurençon H, Saulnier L, Tronchon L, et al. Obelics: An open web-scale filtered dataset of interleaved image-text documents. In: Advances in Neural Information Processing Systems 36 (NeurIPS 2024); 2024 Dec 9-15; Vancouver, BC. Red Hook, NY: Curran Associates, Inc., 2024.

23. HuggingFace. HuggingFace 2024. Available at: https://huggingface.co/ (accessed 6th July 2024).

24. Lin C-Y, Rouge: A package for automatic evaluation of summaries. Text summarization branches out 2004:74-81.

25. Delbrouck J-B, Chambon P, Bluethgen C, et al. Improving the factual correctness of radiology report generation with semantic rewards. arXiv preprint arXiv:2210.12186. 2022.

26. Gu Y, Tinn R, Cheng H, et al. Domain-specific language model pretraining for biomedical natural language processing. ACM Trans Comput Healthc 2021;3:1-23

27. Jain S, Agrawal A, Saporta A, et al. Radgraph: Extracting clinical entities and relations from radiology reports. In: NeurIPS 2021 Datasets and Benchmarks Track; 2021 Dec 6-12; Virtual. Red Hook, NY: Curran Associates, Inc., 2021.

28. Zhang Y, Merck D, Tsai EB, et al. Optimizing the factual correctness of a summary: A study of summarizing radiology reports. In: Proceedings of the 58th Annual Meeting of the Association for Computational Linguistics (ACL); 2020 Jul 5-10; Virtual. Stroudsburg, PA: Association for Computational Linguistics, 2020:5108-5120.

29. Smit A, Jain S, Rajpurkar P, et al. CheXbert: combining automatic labelers and expert annotations for accurate radiology report labeling using BERT. In: Proceedings of the 2020 Conference on Empirical Methods in Natural Language Processing (EMNLP); 2020 Nov 16-20; Virtual. Stroudsburg, PA: Association for Computational Linguistics, 2020:1500-1519.

30. Amazon Web Services I. Amazon EC2 P4 Instances 2024. Available at: https://aws.amazon.com/ko/ec2/instance-types/p4/ (accessed 21st July 2024).

31. Virtanen P, Gommers R, Oliphant TE, et al. SciPy 1.0: fundamental algorithms for scientific computing in Python. Nat Methods 2020;17:261-272.

32. Birkun AA, Gautam A. Large Language Model (LLM)-powered chatbots fail to generate guideline-consistent content on resuscitation and may provide potentially harmful advice. Prehosp Disaster Med 2023;38:757-763.

33. Zack T, Lehman E, Suzgun M, et al. Assessing the potential of GPT-4 to perpetuate racial and gender biases in health care: a model evaluation study. Lancet Digital Health 2024;6:e12-e22.

34. Konstantinov N, Lampert CH. Fairness-aware pac learning from corrupted data. JMLR 2022;23:1-60.

35. Prasser F, Gaupp J, Wan Z, et al., editors. An open source tool for game theoretic health data de-identification. AMIA Annu Symp Proc 2018; 2017:1430–1439.

36. Martín-Noguerol T, López-Úbeda P, Luna A. Large language models in Radiology: The importance of fine-tuning and the fable of the luthier. Eur J Radiol 2024;178:111627.

37. Singhal K, Tu T, Gottweis J, et al. Towards expert-level medical question answering with large language models. arXiv preprint arXiv 2023;2305.09617.

38. Kim H, Hwang H, Lee J, et al. Small language models learn enhanced reasoning skills from medical textbooks. arXiv preprint arXiv 2024;2404.00376.

39. Wang Z, Zhong W, Wang Y, et al. Data management for large language models: A survey. arXiv preprint arXiv 2023;2312.01700.

40. Li J, Li D, Savarese S, et al., editors. Blip-2: Bootstrapping language-image pre-training with frozen image encoders and large language models. In: Proceedings of the 40th International Conference on Machine Learning (ICML); 2023 Jul 23-29; Honolulu, HI. PMLR, 2023:814:1-13.

41.     Liu H, Li C, Wu Q, et al. Visual instruction tuning. In: Advances in Neural Information Processing Systems 36 (NeurIPS 2024); 2024 Dec 9-15; Vancouver, BC. Red Hook, NY: Curran Associates, Inc., 2024.

42.     Lobig F, Subramanian D, Blankenburg M, et al. To pay or not to pay for artificial intelligence applications in radiology. NPJ Digital Medicine. 2023;6:117.

43.     Vargas-Palacios A, Sharma N, Sagoo GS. Cost-effectiveness requirements for implementing artificial intelligence technology in the Women's UK Breast Cancer Screening service. Nat Commun. 2023;14:6110.

44.     Statistics USBoL. Occupational Employment and Wages, May 2023: U.S. Bureau of Labor Statistics; 2023. Available at: https://www.bls.gov/oes/current/oes291224.htm#(5))(accessed 18th July 2024).

45.     Alayrac J-B, Donahue J, Luc P, et al. Flamingo: a visual language model for few-shot learning. In: Advances in Neural Information Processing Systems 35 (NeurIPS 2022); 2022 Nov 28-Dec 9; New Orleans, LA. Red Hook, NY: Curran Associates, Inc., 2022:23716-23736.

**Table 1 Performance Degradation of Multimodal vs. Text-Only LLMs with Increasing Data Corruption**

| Model | Full Finding | Corrupted 0.2 | Corrupted 0.5 | Corrupted 0.8 |
|---|---|---|---|---|
| | ROUGE-L | | | |
| OpenFlamingo | 0.40 | 0.38 | 0.29 | 0.22 |
| OpenFlamingo_llm_only | 0.39 | 0.38 | 0.28 | 0.15 |
| p-value | 0.36 | 0.52 | 0.13 | 1.39E-06 |
| Med-Flamingo | 0.41 | 0.39 | 0.32 | 0.18 |
| Med-Flamingo_llm_only | 0.21 | 0.22 | 0.16 | 0.15 |
| p-value | 5.83E-20 | 6.93E-18 | 1.49E-14 | 0.08 |
| IDEFICS | 0.46 | 0.42 | 0.35 | 0.24 |
| IDEFICS_llm_only | 0.21 | 0.22 | 0.16 | 0.15 |
| p-value | 1.03E-24 | 5.64E-21 | 1.51E-17 | 4.37E-06 |
| | F1RadGraph | | | |
| OpenFlamingo | 0.35 | 0.35 | 0.25 | 0.18 |
| OpenFlamingo_llm_only | 0.34 | 0.35 | 0.24 | 0.13 |
| p-value | 0.35 | 0.84 | 0.16 | 1.6E-04 |
| Med-Flamingo | 0.40 | 0.38 | 0.29 | 0.16 |
| Med-Flamingo_llm_only | 0.17 | 0.17 | 0.12 | 0.12 |
| p-value | 4.44E-20 | 7.25E-21 | 2.75E-15 | 0.03 |
| IDEFICS | 0.44 | 0.39 | 0.31 | 0.21 |
| IDEFICS_llm_only | 0.17 | 0.17 | 0.12 | 0.12 |
| p-value | 2.56E-22 | 2.19E-20 | 1.36E-17 | 1.26E-05 |
| | F1CheXbert | | | |
| OpenFlamingo | 0.51 | 0.49 | 0.41 | 0.36 |
| OpenFlamingo_llm_only | 0.53 | 0.51 | 0.39 | 0.35 |
| p-value | 0.85 | 0.91 | 0.06 | 0.61 |
| Med-Flamingo | 0.61 | 0.57 | 0.46 | 0.36 |
| Med-Flamingo_llm_only | 0.40 | 0.36 | 0.29 | 0.29 |
| p-value | 2.05E-05 | 1.86E-06 | 0.0027 | 0.33 |
| IDEFICS | 0.62 | 0.57 | 0.47 | 0.38 |
| IDEFICS_llm_only | 0.40 | 0.36 | 0.29 | 0.29 |
| p-value | 1.80E-06 | 1.28E-07 | 9.37E-06 | 0.06 |

The table compares the efficacy of multimodal and text-only LLMs across standard and corrupted data scenarios. It succinctly presents the models' performance degradation with increasing data corruption and underscores the statistical significance of these differences, reflecting the added value of multimodal learning. Abbreviations: 'OpenFlamingo' is the 3B parameter model; 'OpenFlamingo_llm_only' excludes vision. 'Med-Flamingo' and 'IDEFICS' are 9B parameter models; their 'llm_only' versions are text-only.

**Table 2. Performance Comparison of Fine-Tuned vs. General Models on Standard and Corrupted Data**

| Model | Full Finding | Corrupted 0.2 | Corrupted 0.5 | Corrupted 0.8 |
|---|---|---|---|---|
| | ROUGE-L | | | |
| OpenFlamingo | 0.40 | 0.38 | 0.29 | 0.22 |
| Med-Flamingo | 0.414 | 0.39 | 0.32 | 0.18 |
| p-value | 0.19 | 0.10 | 4.30E-4 | 0.18 |
| | F1RadGraph | | | |
| OpenFlamingo | 0.35 | 0.35 | 0.25 | 0.18 |
| Med-Flamingo | 0.40 | 0.38 | 0.29 | 0.16 |
| p-value | 5.07E-04 | 4.59E-03 | 2.23E-03 | 0.35 |
| | F1CheXbert | | | |
| OpenFlamingo | 0.51 | 0.49 | 0.41 | 0.36 |
| Med-Flamingo | 0.61 | 0.57 | 0.46 | 0.36 |
| p-value | 1.04E-01 | 0.13 | 0.62 | 0.04 |

Abbreviations: 'OpenFlamingo' is the 3B parameter model; 'Med-Flamingo' is a 9B parameter model.

**Figure Legends**

Fig. 1. Three examples of randomly masked subwords at rates of 20%, 50%, and 80%. Sub-words divide words into smaller units. In the field of natural language processing, the most commonly used method for training language models is the sub-word approach. Randomly selected sub-words are replaced with underscores '_'.

| Full | Corrupted 0.2 | Corrupted 0.5 | Corrupted 0.8 |
|---|---|---|---|
| mild to moderate enlargement of the cardiac silhouette is unchanged. the aorta is calcified and diffusely tortuous. the mediastinal and hilar contours are otherwise similar in appearance. there is minimal upper zone vascular redistribution without overt pulmonary edema. no focal consolidation, pleural effusion or pneumothorax is present. the osseous structures are diffusely demineralized | mild to moderate enlargement _ _ cardiac silhouette is unchanged. _ aorta is calcified and _ _ tortuous. the media _nal and hilar contours are otherwise _ _ _ appearance. _ is minimal upper zone _ redistribution without overt pulmonary edema _ no _ consolidation, _eural e _ _ or pneum _rax _ present. the osseous structures are diffusely demi _alized | _ to moderate en _ _ _ _ _ silhouette _ unchanged _ the ao _ is calci _ _ diffusely tortu _. _ _ _ _ and _ _ _ _ _ are _ similar _ appearance. _ is _ upper _ _ vascular redistribution _ _ _ pulmonary _ _ _ _ focal _, pleural _ _ _ or pneumothorax is present _ the _se _ are _ly demi _ _zed _ | _ _ moderate _ _ _ _ _ cardiac _ _ unchanged _ the _rta _ _cified and _ _ _ _ _ _ _ _ _ _ _ _ _ _ _ _ _ _ _ _ _ _ upper _ _ _ _ _ _ ed _ _ no _ _eur _ e _ _ or _ _ _ otho _ _ _ _ _ the _ _ _ _ are _ly _ _ _ _ _ |
| frontal and lateral views of the chest were obtained. there is mild basilar atelectasis without evidence of focal consolidation. no pleural effusion or pneumothorax is seen. there is minimal biapical pleural thickening. cardiac silhouette is top normal with likely adjacent epicardial fat pad. the aorta is calcified and tortuous. some degenerative changes are seen along the spine | _ and _ views of the _ were obtained. _ is _ basilar ate _ _sis without evidence of focal consolidation. no pleural effusion or pneumotho _x is _. there _ minimal biapical pleural thickening. cardiac silhouette is top normal with likely adjacent epic _ial fat pad. the ao _ _ _ci _ and _tuous. _ degenerative changes are seen _ the spine | frontal _ _ _ of the _ _ obtained. _ is _ basilar _lectasis _ _ _ _ _ _ _ _ _ _ _al e _ _ _ pneumothora _ is seen. there _ minimal _ _ _ _eural thick _. cardiac _ is _ _ with likely adjacent epicard _ fat pad. _ aorta _ cal _ _ _ _tu _. _ _ _ _ _ _ seen along the spine _ | _ _ _ views _ _ _ _ _ _ _ _ _ar ate _cta _ _ _ _ _ _. _ _ _ _ _ _ _ _ _ _ _ _ is seen _ _ _ minimal bi _ _ _ _al _ening _ _ _ _ _ _ _ likely _ _ _ial _ pad _ the ao _ _ _ci _ _ _ _ _ _ _ _ changes _ _ _ the _ _. |
| heart size is mildly enlarged. there is mild unfolding of the thoracic aorta. cardiomediastinal silhouette and hilar contours are otherwise unremarkable. there is mild bibasilar atelectasis. lungs are otherwise clear. pleural surfaces are clear without effusion or pneumothorax. focus of air seen under the right hemidiaphragm, likely represents colonic interposition | heart size _ mildly enlarged. there is mild un _ of the thorac _ aorta. _iomediastinal silhouette and _lar con _ _s are otherwise _remarkable. there is mild bibasila _ ate _cta _. lungs are otherwise clear. pleur _ surfaces _ clear without effusion _ pneum _rax _ focus of _ seen under the right hemidiaphragm, _ represents colonic inter _. | heart _ _ _ enlarged _ there _ mild _folding of the thoracic ao _ _ _ cardiomed _ _l silhouette and hi _ _ contours _ otherwise _ _ _able. _ is _ bi _ _ _ _ _ctasis _ _ are _ clear. pl _ _ _ _ clear _ _ffusion _ pneumotho _ _ _ focus _ air _ _ _ _ hem _ph _m _ _ represents _ic _ _. | _ _ _ _ _ _ _ _ _ folding _ _ _ _ _ _ _. card _ _ _ _l silhouette _ _ _ con _ _ _ _ _ _ _ _ _ _ _ _ _ ila _ _lecta _. _ _ _ _ _ _ _eural _ _ _ _ _ _ _ _ _ _ othora _ _ _ _ _ _ seen _ the _ _ _ _ _m _ likely _ _ _ _ _. |

**Fig. 2. LLM workflow with performance metrics evaluation. The entire workflow is based on six large language models (LLMs) (three multimodal and three unimodal). Each model generates impressions from various input data. The performance is evaluated using the ROUGE-L, F1RadGraph, and F1CheXbert metrics.**

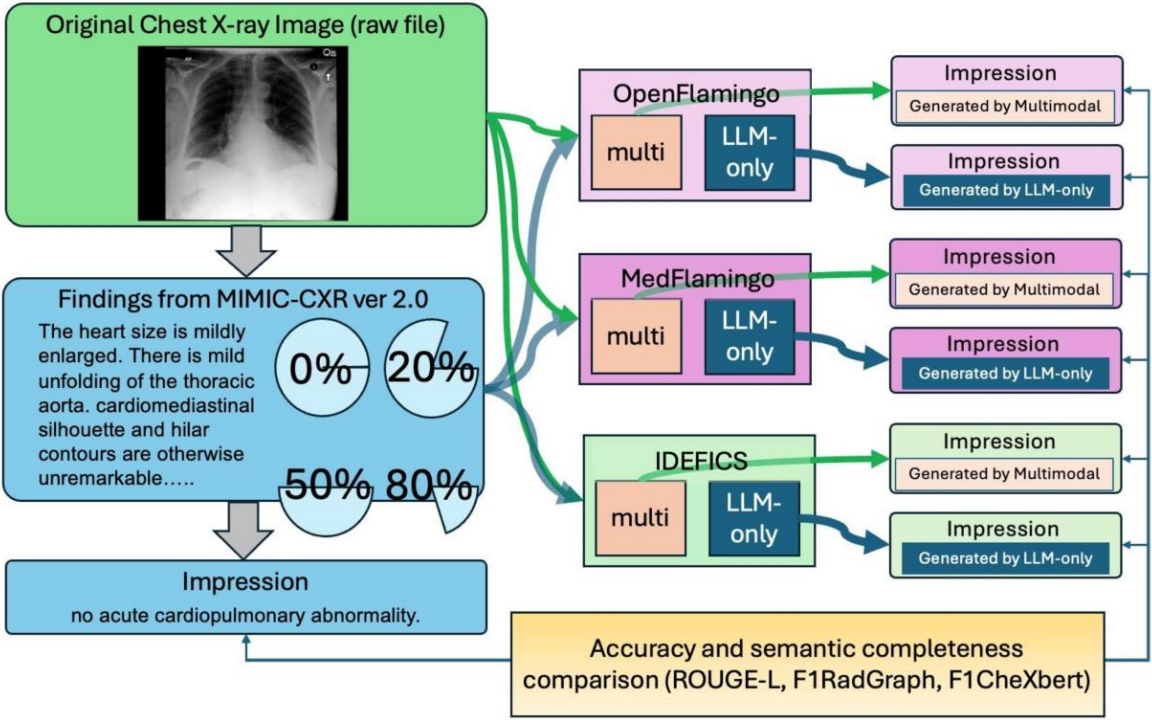

Fig. 3. Performance degradation of multimodal vs. text-only LLMs with data corruption. This graph compares full models with their text-only versions across ROUGE-L (Graph A), F1RadGraph (Graph B), and F1CheXbert (Graph C) metrics, showing performance on clean and variously corrupted data levels (0.2, 0.5, and 0.8). Statistical significance is indicated by asterisks: *** ($p < 0.001$) signifies a highly significant difference, ** ($p < 0.01$) indicates a moderately significant difference, and * ($p < 0.05$) denotes a nominally significant difference.

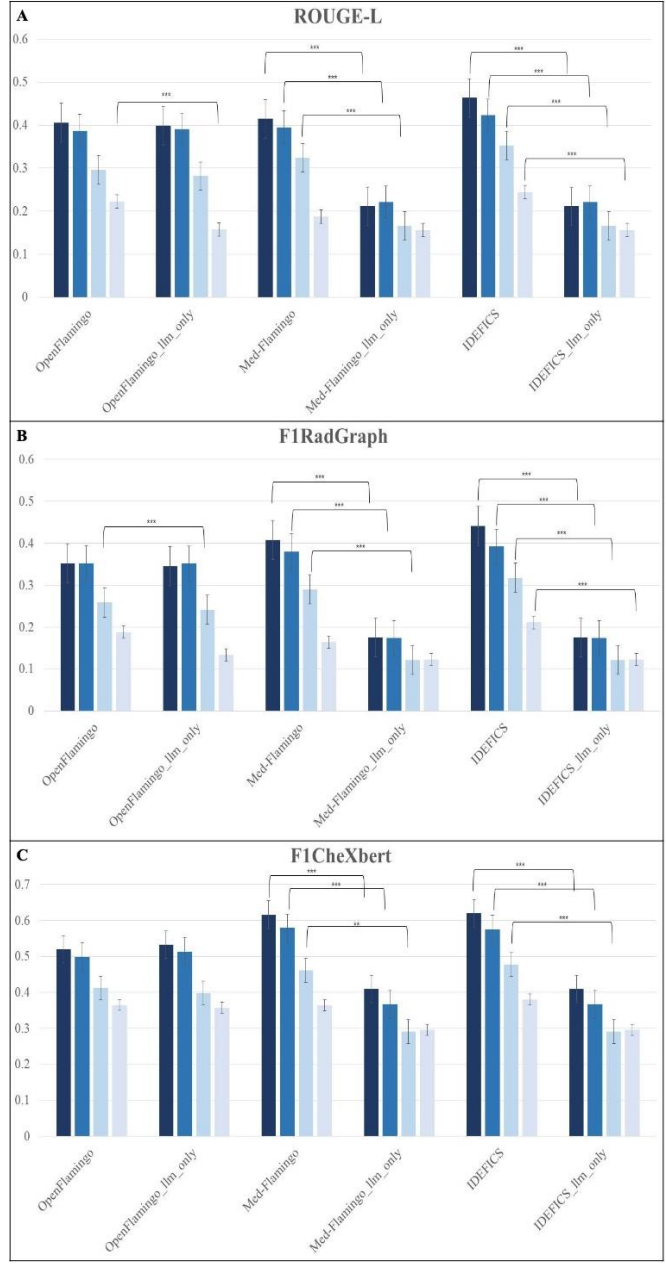

Fig. 4. Examples of model predictions compared to radiologist's reference: outputs by MedFlamingo (multimodal) for five representative diseases (both correct and incorrect answers were displayed). The target labels were classified by CheXbert. Correct answers were highlighted in blue, while incorrect answers were highlighted in red. The findings were presented in full, without any corruption.

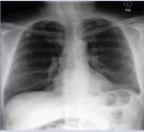